\documentclass[12pt]{iopart}

\begin{document}
\title[Free particle on SU(2) manifold]{Free particle on $SU(2)$ group
manifold}
\author{George Chavchanidze, Levan Tskipuri}
\address{Department of Theoretical Physics, A. Razmadze Institute of
Mathematics,\\
1 Aleksidze Street, Ge 380093, Tbilisi, Georgia\\
e-mails: chaffinch@apexmail.com, tskipa@usa.net}

\begin{abstract}
We consider classical and quantum dynamics of a free particle on 
SU(2) group manifold.
The eigenfunctions of the Hamiltonian are constructed 
in terms of coordinate free objects
\end{abstract}

\section{Lagrangian description}

The dynamics of a free particle on $SU(2)$ group manifold is described
by the Lagrangian 
\begin{equation}{\cal L}=\langle g^{-1}\, \dot g\, g^{-1}\,\dot g \rangle
\end{equation}
where $g\in SU(2)$ and $\langle ~ \rangle$ denotes the normalized trace
$\langle ~\rangle=-\frac{1}{2}Tr(~)$, which defines a scalar product in
$su(2)$.
The Lagrangian (1.1)
defines the following dynamical equations 
\begin{equation}
\frac{d}{dt}\langle g^{-1}\dot g\rangle=0
\end{equation}
otherwise ,one can notice that our Lagrangian has $SU(2)$ "right"
and $SU(2)$ "left" symmetry .It means that it is invariant under the
following
transformations
\begin{equation}
g\longrightarrow h_1\,g~~~~~~~~~~~~~~\mbox {"left" symmetry}      
\end{equation}
\begin{equation}
~g\longrightarrow g\,h_2~~~~~~~~~~~~~~\mbox {"right" symmetry}
\end{equation}
According to the Noether's theorem the corresponding conserving quantities
are
\begin{equation}
~R=g^{-1}\dot g ~~~~~~~\frac{d}{dt}R=0~~~~~~~~~~~~~\mbox  {"right"
symmetry}
\end{equation}
\begin{equation}
L= \dot g\,g^{-1}~~~~~~~ \frac{d}{dt}L=0~~~~~~~~~~~~\mbox{"left" symmetry}
\end{equation}
Now let's introduce the basis of $su(2)$ algebra
\begin{equation}
T_1=\left( \begin{array}{cr}
i&0\\0&-i \end{array}\right),~~~~
T_2=\left( \begin{array}{cr}
0&~1\\-1&~0 \end{array}\right),~~~~
T_3=\left( \begin{array}{cr}
0&i\\i&0 \end{array}\right),
\end{equation}
 The elements of $su(2)$ are traceless anti-hermitian matrices, and any 
$A\in su(2)$ can be parameterized in the following way
\begin{equation}
A= A^nT_n~~~~~~~~~n=1,2,3~~~~~~~~~~R= R^nT_n
\end{equation}
 Scalar product $AB=\langle AB\rangle=-\frac{1}{2}Tr(AB)$ provides that 
\begin{equation}
A^n=\langle A\,T_n\rangle~~~~~~~~~~~(\langle T_nT_m\rangle=\delta _{nm})
\end{equation}
Now we can introduce 6 functions
\begin{equation}
R_n=\langle T_nR \rangle~~~~~~~~~n=1,2,3~~~~~~~~~R=R^nT_n
\end{equation}
\begin{equation}
~L_n=\langle T_nL \rangle~~~~~~~~~n=1,2,3~~~~~~~~~L=L^nT_n
\end{equation}
which are integrals of motion.
It is easy to find general solution of Euler-Lagrange equation
\begin{equation}
\frac{d}{dt}g^{-1}\dot g=0~~~~~~~~\Rightarrow~~~~~~~g^{-1}\dot g=const
\end{equation}
\begin{equation}
g=e^{Rt}g(0)
\end{equation}
These are well known geodesics on Lie group.

\section{Hamiltonian description}

Working in a first order Hamiltonian formalism we construct new Lagrangian 
which is equivalent to the initial one
\begin{equation}
\tilde {\cal L}=\langle R(g^{-1} \dot g-\tilde v)\rangle +\frac{1}{2}
\langle \tilde v^2\rangle
\end{equation}
in sense that variation of R provides
\begin{equation}
 g^{-1} \dot g =\tilde v
\end{equation}
and $ \tilde {\cal L} $ reduces to $ \cal{L} $.
Variation of $\tilde v $ gives $R=\tilde v $ and therefore we can rewrite 
equivalent Lagrangian $\tilde {\cal L} $ in terms of R and g
\begin{equation}
\tilde {\cal L}=\langle R\,g^{-1} \dot g \rangle -\frac{1}{2} \langle R^2
\rangle
\end{equation}
Where function $\frac{1}{2} \langle R^2 \rangle $ plays role of
Hamiltonian and 
one-form $ \langle R\,g^{-1}\,d\,g \rangle $ is a symplectic potential
$\theta $. 
External differential of  $ \theta $ is the symplectic form $ \omega
$,that determines
Poisson brackets and the form of Hamilton's equation.
\begin{equation}
\omega =d\,\theta =-\langle g^{-1}d\,g\,\wedge d\,R \rangle -\langle
R\,g^{-1} d\,g\,\wedge \, g^{-1}d\,g \rangle
\end{equation}
$\omega $ provides isomorphism between vector fields and one-forms 
\begin{equation}
X~~\rightarrow ~~~X\rfloor \omega
\end{equation}
Let ${\cal F}(SU(2))$ denote the real-valued smooth function on
$SU(2)$.For an $f\in SU(2)$ there exists a Hamiltonian vector field
satisfying  
\begin{equation}
X_f:~~~~~~~ X\rfloor \omega =-df
\end{equation}
Where $ X\rfloor \omega $ denotes the contraction of $X$ with $\omega $.  
$X_f$ is called Hamiltonian vector field associated with $f$. According to
the definition Poisson bracket of two function is as follows
\begin{equation}
\{f,g\} = {{\cal L}_X}_f g=X_f\rfloor dg =\omega(X_f,X_g)
\end{equation}
Where ${{\cal L}_X}_f g$ denotes Lie derivative of $g$ with respect to
$X_f$.
The skew symmetry of $\omega $ provides skew symmetry of Poisson bracket .
Hamiltonian vector fields that correspond to $R_n,L_m $ and $g $ functions
are the following
\begin{equation}
X_n=X_{R_n}=(~[R,T_n]~,~g\,T_n~)=(~X_{R_n}^{(R)}~,X_{R_n}^{(g)})
\end{equation}
\begin{equation}
Y_n=X_{L_m}=(~[R,g\,T_mg^{-1}]~,~T_mg~)=(~X_{L_m}^{(R)}~,X_{L_m}^{(g)})
\end{equation}
Therefore Poisson brackets are 
\begin{equation}
\{ L_n,L_m\} =-2\epsilon_{nm}^{~~~k}\,L_k
~~~\{ R_n,R_m\} =2\epsilon_{nm}^{~~~k}\,R_k
\end{equation}
\begin{equation}
\{ R_n,L_m\} =0
\end{equation}
\begin{equation}
\{ R_n,g\} = g\,T_n
~~~~~\{ L_m,g\} = T_mg
\end{equation}
 the results are natural. $R$ and $L$ that correspond respectively to the
"right" 
and "left" symmetry commute with each other and independently form $su(2)$ 
algebras.
    It is easy to write down Hamilton's equations
\begin{equation}
\dot g=\{ H,\,g\}=g\,R
\end{equation}
\begin{equation}
\dot R=\{ H,\,R\}=0
\end{equation}
 We consider case of $SU(2)$ , but the same constructions can be applied
to
the other Lie groups.

\section{Quantization}

Let's introduce operators 
\begin{equation}
\hat R_n =~\frac{i}{2}{{\cal L}_X}_n~~
\end{equation}
\begin{equation}
\hat L_m =-\frac{i}{2}{{\cal L}_Y}_m
\end{equation}
They act on the square integrable functions(see Appendix A) on $SU(2)$ and
satisfy quantum
commutation relations
\begin{equation}
[{\hat L}_n,{\hat L}_m] =i\epsilon_{nm}^{~~~k}\,{\hat L}_k
\end{equation}
\begin{equation}
[{\hat R}_n,{\hat R}_m] =i\epsilon_{nm}^{~~~k}\,{\hat R}_k
\end{equation}
\begin{equation}
[ {\hat R}_n,{\hat L}_m] =0~~~~~~~~
\end{equation}
  The Hamiltonian is defined as 
\begin{equation}
\hat H=\hat R^2=\hat L^2
\end{equation}
Therefore the complete set of observables that commute with each other
is as follows
\begin{equation}
\hat H,~~~~\hat R_a,~~~~~\hat L_b
\end{equation}
Where a and b unlike n and m are fixed.Using a simple generalization of a
well known algebraic construction (see Appendix B)one can check that the
eigenvalues of the quantum observables
 $ \hat H $,$ {\hat R}_a $ and $ {\hat L}_b $ are as follows
\begin{equation}
\hat H~\psi_{lr}^j=j(j+1)~\psi_{lr}^j
\end{equation}
where $j$ takes positive integer and half integer values
\begin{equation}
j=0,~~\frac{1}{2},~~1,~~\frac{3}{2},~~2~...
\end{equation}
\begin{equation}
\hat R_a~\psi_{lr}^j=r~\psi_{lr}^j
\end{equation}
\begin{equation}
\hat L_b~\psi_{lr}^j =l~\psi_{lr}^j
\end{equation}
 with $r$ and $l$ taking values in the following range
\begin{equation}
-j,~~-j+1,~...~,~~j-1,~~j
\end{equation}
The main aim of the article is construction of the corresponding
eigenfunctions
$~\psi_{lr}^j$.The first step of this construction is proposition 1 \\
{\bf proposition 1.}
The function $\langle \tilde T\,g \rangle $ where $\tilde
T=(I+iT_a)(I+iT_b)$
is an eigenfunction of $\hat H$,${~~\hat R}_a$ and ${\hat L}_b $ with
eigenvalues,
respectively $\frac{3}{4},~\frac{1}{2}$ and $\frac{1}{2}$

Proof of this proposition is straightforward. Using  $\langle \tilde T\,g
\rangle$ 
we construct the complete set of eigenfunctions of $\hat H$,$~~{\hat R}_a$ 
and ${\hat L}_b$ operators 
\label{ eigenfunctions}
\begin{equation}
~\psi_{lr}^j={\hat L}_-^{j-l}{\hat R}_-^{j-r}~{\langle \tilde T\,g \rangle
}^{2j} 
\end{equation} 
(for the definition of the ${\hat R}_- $ and ${\hat L}_- $ operators see
Appendix)
that are defined  up to a constant multiple. Indeed , acting on
(40) with $\hat H$,${~~\hat R}_a$ and ${\hat L}_b$ operators
 and using commutation relations (see Appendix B) one can prove that
equations
(35-38) hold for $~\psi_{lr}^j$ defined by (40)\\

\section{Free particle on $S^2$ as a $SU(2)/U(1)$ coset model}

  Free particle on $2D$ sphere can be obtained from our model by gauging
$U(1)$ symmetry. In other words let's consider the following local gauge
transformations
\begin{equation}
g \longrightarrow	h(t)\,g  \label{eq:gauge}
\end{equation}
  Where $h(t)\in U(1) \subset SU(2)$ is an element of $U(1).$ Without loss
of generality we can take 
\begin{equation}
h=e^{\,\beta (t)\,T_3}
\end{equation}
  Since $T_3$ is antihermitian $h(t) \in U(1)$ and since $h(t)$ depends on
$t$ Lagrangian 
\begin{equation}
{\cal L}=\langle g^{-1}\, \dot g\, g^{-1}\,\dot g \rangle \label{eq:lag}
\end{equation}
 is not invariant under (\ref{eq:gauge}) local gauge transformations.
  To make (\ref{eq:lag}) gauge invariant we should replace $\frac{d}{dt}$
with a covariant 
  derivative $\nabla g=(\frac{d}{dt}+B)\,g$
  Where $B$ can be represented as follows
\begin{equation}
B=b\,T_3\in su(2)
\end{equation}
with transformation rule
\begin{equation}
B \longrightarrow h\,B\,h^{-1}-\dot h\,h^{-1}
\end{equation}
in the other words
\begin{equation}
b \longrightarrow b- \dot \beta 
\end{equation}
The new Lagrangian
\begin{equation}
{\cal L}_G=\langle g^{-1}\, \nabla g\, g^{-1}\,\nabla g \rangle
\label{eq:newlag}
\end{equation}
is invariant under (\ref{eq:gauge}) local gauge transformations. But this 
Lagrangian as well as every gauge invariant Lagrangian is singular. 
It contains additional non-physical degrees of freedom. To 
eliminate them we should eliminate $B$ using Lagrange equations
\begin{equation}
\frac{\partial {\cal L}_G}{\partial B}=0~~~~ \Longrightarrow
~~~~b=-\langle 
\dot g\,g^{-1}\,T_3 \rangle
\end{equation} 
put it  back in (\ref{eq:newlag}) and rewrite last obtained Lagrangian in
terms of gauge invariant (physical) variables.
\begin{equation}
{\cal L}_G=\langle (g^{-1}\, \dot g\,-L_3T_3)^2 \rangle
\end{equation}
It's obvious that the following 
\begin{equation}
X=g^{-1}T_3\,g~\in ~su(2)
\end{equation}
element of $su(2)$ algebra is gauge invariant. Since $X \in su(2)$ it can
be parameterized as follows 
\begin{equation}
X=x^a\,T_a
\end{equation} 
where $x^a$ are real functions on $SU(2)$ 
\begin{equation}
x_a=\langle X\,T_a \rangle
\end{equation}
  So we have three gauge invariant variables $~~x^a~~~a=1,2,3$ but it's
easy to 
check that only two of them are independent. Indeed
\begin{equation}
\langle X^2\rangle =\langle g^{-1}T_3\,g\,g^{-1}T_3\,g\rangle =\langle
T_3^2\rangle =1
\end{equation}
otherwise
\begin{equation}
\langle X^2\rangle =\langle x^a\,T_a\,x^b\,T_b\rangle =x^a\,x_a
\end{equation}
  So physical variables take values on $2D$ sphere. In other words
configuration space of $SU(2)/U(1)$ model is sphere.
  By direct calculations one can check that having 
  been rewritten in terms of gauge invariant variables ${\cal L}_G$ 
  takes the form 
\begin{equation}
{\cal L}_G=\frac{1}{4}\langle X^{-1}\, \dot X\, X^{-1}\,\dot X \rangle
\end{equation}
  This Lagrangian describes free particle on the sphere. Indeed, 
  since $X=x^a\,T_a$ it's easy to show that 
\begin{equation}
{\cal L}_G=\frac{1}{4}\langle X^{-1}\, \dot X\, X^{-1}\,\dot X \rangle =
\frac{1}{4}\langle X\, \dot X\, X\,\dot X \rangle =\frac{1}{2}\dot
x^a\,\dot x_a
\end{equation}
  So $SU(2)/U(1)$ coset model describes free particle on $S^2$ manifold.

\section{ Quantization of the coset model.}
  
Working in a first order Hamiltonian formalism (see (14)-(16))we get 
\begin{equation}
\tilde {\cal L}_G=\langle R(g^{-1} \dot g-\tilde u)\rangle +\frac{1}{2} 
\langle (\tilde u+g^{-1}\,B\,g)^2\rangle
\end{equation}
\begin{equation}
\tilde {\cal L}=\langle R\,g^{-1} \dot g \rangle -\frac{1}{2} \langle R^2
\rangle
\end{equation}
variation of $\tilde u$ provides:
\begin{equation}
R=\tilde u + g^{-1}\,B\,g 
\end{equation}
\begin{equation}
\tilde u=R-g^{-1}\,B\,g
\end{equation}
  Rewriting $\tilde {\cal L}_G $ in terms of $R$ and $g$ leads to 
\begin{eqnarray}
\tilde {\cal L}_G=\langle R\,g^{-1} \dot g \rangle -\frac{1}{2} \langle
R^2
 \rangle -\langle B\,g\,R\,g^{-1}\rangle =\langle R\,g^{-1} \dot g \rangle 
 -\frac{1}{2} \langle R^2 \rangle -b\langle g\,R\,g^{-1}\,T_3\rangle
\nonumber \\
=\langle R\,g^{-1} \dot g \rangle -\frac{1}{2} \langle R^2 \rangle -b\,L_3
\end{eqnarray}
 Due to the gauge invariance of (\ref{eq:newlag}) we obtain constrained
Hamiltonian system, 
 where $\langle R\,g^{-1}\,dg\rangle $ is symplectic potential, 
 $\frac{1}{2}\langle R^2\rangle$ plays role of Hamiltonian and 
 $b$ is a Lagrange multiple, variation of which leads to the first class 
 constrain:
\begin{equation}
\phi =\langle g\,R\,g^{-1}\,T_3\rangle =\langle L\,T_3\rangle =L_3=0
\label{eq:constr}
\end{equation}
Therefore coset model is equivalent to the initial one with
(\ref{eq:constr}) constrain.
  Using technique of the constrained quantization, instead of 
  quantization of the coset model we can submit quantum model ,that 
  corresponds to the free particle on $SU(2)$, to the following 
  operator constrain 
\begin{equation}
{\hat L}_3|\psi \rangle =0
\end{equation}
                                             
$~$Free particle on $SU(2) ~~~~~\longrightarrow  ~~~~$ Quantum particle on
$SU(2)$
 
reduction $~~\downarrow
~~~~~~~~~~~~~~~~~~~~~~~~~~~~~~~~~~~~~~~~~~~~~~~~~~\downarrow ~
~~ $reduction
                                                  
$~$Free particle on $S^2~~~~~~~~~~\longrightarrow $ ~~~~~Quantum particle
on $S^2$
\\
\\
  Hilbert space of the initial sistem (that is linear span of $\psi
_{rl}^j
~~~~j=0,~\frac{1}{2},~1,~\frac{3}{2},~2,~...$ wave functions) reduces to 
the linear span of $\psi _{r0}^j~~~~j=0,~1,~2,~3,~...$  wave functions.
Indeed, 
${\hat L}_3\psi _{rl}^j$  implies $l=0$, and since $l=0~~~j$ is integer. 
Therefore $r$ takes $-j,~-j+1,~...,~j-1,~j$ integer values only.
  Wave functions $\psi _{rl}^j$ rewriten in terms of gauge invariant 
  variables  up to a constant multiple should coincide with well known 
  spherical harmonics
\begin{equation}
\psi _{r0}^j \sim {\cal J}_{jr}
\end{equation}
One can chack the following 
\begin{equation}
\psi _{r0}^j \sim {\hat L}_-^j{\hat R}_-^{j-r}~{\langle \tilde T\,g
\rangle }^{2j} \sim {\hat R}_-^{j-r}~{\langle T_+\,g^{-1}\,T_3\,g \rangle
}^j \sim 
 {\hat R}_-^{j-r}\,sin^j\,\theta \,e^{ij\theta } \sim {\hat
R}_-^{j-r}\,{\cal J}_{jj}
\sim {\cal J}_{jr}
\end{equation}

  This is an example of using large initial model in quantization of 
  coset model.
\\

\section{ Appendix A}
\def\theequation{A.\arabic{equation}}

Scalar  product in Hilbert space is defined as  follows
\begin{equation}
\langle \psi _1 | \psi _2 \rangle =\int _{SU(2)} \prod _{a=1}^{3}
\langle g^{-1}\,dg\,T_a \rangle \psi _1^*  \,\psi _2 \label{eq:scal}
\end{equation}
It's easy to prove that if scalar product is (\ref{eq:scal})  operators 
${\hat R}_n $ and $ {\hat L}_m $ are  hermitian.
Indeed 
\begin{equation}
\langle \psi _1 |{\hat R}_n \psi _2\rangle = \int _{SU(2)} \prod
_{a=1}^{3}
\langle g^{-1}\,dg\,T_a \rangle \psi _1^*  \,(\frac{i}{2}{{\cal L}_X}_n 
\psi _2) \label{scal}=\int _{SU(2)} \prod _{a=1}^{3}
\langle g^{-1}\,dg\,T_a \rangle (\frac{i}{2}{{\cal L}_X}_n\psi _1)^*  
\,\psi _2 \label{scal}
\end{equation}
Where integration by part have been used. It's easy to check that the
additional term coming from measure 
\begin{equation}
\prod _{a=1}^{3}\langle g^{-1}\,dg\,T_a \rangle
\end{equation}
vanishes since 
\begin{equation}
{{\cal L}_X}_n \langle g^{-1}\,dg\,T_a \rangle
\end{equation}
  For more transparency one can introduce the following parameterization
of 
  $SU(2)$.  For any $g \in SU(2)$.
\begin{equation}
g=e^{q^a\,T_a}
\end{equation}    
  Then the symplectic potential takes the form 
\begin{equation}
\langle R\,g^{-1}\,dg \rangle =R_a\,dq^a
\end{equation} 
and scalar product 
\begin{equation}
\langle \psi _1 |\psi _2\rangle =\int _0^{2\pi }\int _0^{2\pi }\int
_0^{2\pi }
d^3q\, \psi _1^*  \,\psi _2
\end{equation}
 that coincides with (\ref{eq:scal}) because of 
\begin{equation}
dq_a=\langle g^{-1}\,dg \,T_a\rangle
\end{equation}
\section{ Appendix B}
\def\theequation{B.\arabic{equation}}

Without loss of generality we can take $\hat H$,$~\hat L_3 $ and $\hat
R_3$ as a complete set of observables. Assuming that there exist at least
one eigenfunctions of $\hat H$,$\hat L_3 $and $\hat R_3$ operators: 
\begin{equation}
\hat H\psi =E\psi 
\end{equation}
\begin{equation}
\hat R_3 \psi =r\psi
\end{equation}
\begin{equation}
\hat L_3\psi =l\psi
\end{equation}
It is easy to show that  
eigenvalues of $\hat H$ are non-negative 
\begin{equation}
 E \geq 0 
\end{equation}
and
\begin{equation}
E-r^2\geq 0  \label{eq:restr1}
\end{equation}
\begin{equation}
E-l^2\geq 0  \label{eq:restr2}
\end{equation}
Indeed operators $\hat R$ and $\hat L $ are selfadjoint so 
\begin{equation}
\langle \psi |\hat H |\psi \rangle =\langle \psi |{\hat R}^2 |\psi \rangle
=\langle \psi |{\hat R}_a{\hat R}^a |\psi \rangle =\langle \psi |{{\hat
R}_a}^{\dagger }{\hat R}^a |\psi \rangle =
\langle {\hat R}_a\psi |{\hat R}^a\psi \rangle =\parallel
{\hat R}_a \psi \parallel \geq 0
\end{equation}

 To prove (\ref{eq:restr1})-(\ref{eq:restr2}) we shall consider ${\hat
R}_1^2+{\hat R}_2^2$ and
${\hat L}_1^2+{\hat L}_2^2$ operators 
\begin{equation}
\langle \psi |{\hat R}_1^2+{\hat R}_2^2 |\psi \rangle =\parallel
{\hat R}_1 \psi \parallel + \parallel {\hat R}_2 \psi \parallel \geq 0
\end{equation}
and
\begin{equation}
\langle \psi |{\hat R}_1^2+{\hat R}_2^2|\psi \rangle =\langle \psi |\hat
H-{\hat R}_3^2 |\psi \rangle =(E-r^2)\langle \psi |\psi \rangle
\end{equation}
Therefore $E-r^2 \geq 0$
Now let's introduce new operators 
\begin{equation}
{\hat R}_+=i{\hat R}_1+{\hat R}_2~~~~~~~~~{\hat R}_-=i{\hat R}_1-{\hat
R}_2
\end{equation}
\begin{equation}
{\hat L}_+=i{\hat L}_1+{\hat L}_2~~~~~~~~~{\hat L}_-=i{\hat L}_1-{\hat
L}_2
\end{equation}
These operators are not selfadjoint, but ${\hat R}_-^{\dagger}={\hat R}_+$
and 
${\hat L}_-^{\dagger}={\hat L}_+$
and they fulfill the following commutation relations 
\begin{equation}
[{\hat R}_{\pm},{\hat R}_3]=\pm {\hat R}_{\pm}
\end{equation}
\begin{equation}
[{\hat L}_{\pm},{\hat L}_3]=\pm {\hat L}_{\pm}
\end{equation}
\begin{equation}
[{\hat R}_+,{\hat R}_-]=2{\hat R}_3
\end{equation}
\begin{equation}
[{\hat L}_+,{\hat L}_-]=2{\hat L}_3
\end{equation}
\begin{equation}
[{\hat R}_*,{\hat L}_*]=0~~~
\end{equation}
where * takes values +,--,3 using these commutation relations it is easy
to show 
that if $ \psi_{rl}^\lambda $ is eigenfunction of $\hat H$,${\hat L}_3$
and 
${\hat R}_3$ with corresponding eigenvalues :
\begin{equation}
\hat H\psi _{rl}^{\lambda }=\lambda \psi _{rl}^{\lambda }
\end{equation}
\begin{equation}
{\hat L}_3{\psi _{rl}^{\lambda }}=l{\psi _{rl}^{\lambda }}
\end{equation}
\begin{equation}
{\hat R}_3{\psi _{rl}^{\lambda }}=r{\psi _{rl}^{\lambda }}
\end{equation}
then ${\hat R}_{\pm}\psi _{rl}^{\lambda } $ and ${\hat L}_{\pm}\psi
_{rl}^{\lambda } $ 
are the eigenfunctions with corresponding eigenvalues
$\lambda ,l\pm 1,r$ and $\lambda ,l,r\pm 1$.
Consequently using ${\hat R}_{\pm },{\hat L}_{\pm }$ operators we
construct 
a family of eigenfunctions with eigenvalues 
\begin{equation}
l,~~~l \pm 1,~~~l \pm 2,~~~l \pm 3,~~~...     \label{eq:seq1}
\end{equation}
\begin{equation}
r,~~~r \pm 1,~~~r \pm 2,~~~r \pm 3,~~~...     \label{eq:seq2}
\end{equation}
but conditions (\ref{eq:restr1}) and (\ref{eq:restr2}) give restrictions
on  a possible range of eigenvalues. 
We should have 
\begin{equation}
\lambda -r^2 \geq 0
\end{equation}
\begin{equation}
\lambda -l^2 \geq 0
\end{equation}

In other words, in order to interrupt (\ref{eq:seq1})-(\ref{eq:seq2})
sequences we should have 
\begin{equation}
{\hat L}_+ {\psi _{rj}^{\lambda }}=0~~~~{\hat L}_-{\psi _{r,-j}^{\lambda
}}=0
\end{equation}
\begin{equation}
{\hat R}_+{\psi _{kl}^{\lambda }}=0~~~~{\hat R}_-{\psi _{-kl}^{\lambda
}}=0
\end{equation}
and for some $j$ and $k$
therefore $l$ and $r$ take the following values 
\begin{equation}
-j,~~~-j+1,~~~...~~j-1~~~,j
\end{equation}
\begin{equation}
-k,~~~-k+1,~~~...~~k-1~~~,k
\end{equation}
The number of values is $2j+1$ and $2k+1$ respectively. Since number of
values 
should be integer , $j$ and $k$ should take integer or half integer values
\begin{equation}
j=0,~~~\frac{1}{2},~~~1,~~~\frac{3}{2},~~~2,~~~...
\end{equation}
\begin{equation}
k=0,~~~\frac{1}{2},~~~1,~~~\frac{3}{2},~~~2,~~~...
\end{equation}
Now using commutation relations we can rewrite $\hat H$ in terms of
$ {\hat R}_{\pm},{\hat R}_3 $ operators:
\begin{equation}
\hat H={\hat R}_+ {\hat R}_-+{\hat R}_3^2+{\hat R}_3   \label{eq:ham}
\end{equation}
( \ref{eq:ham} ) provides that $\lambda =j(j+1)=k(k+1)$
so $j=k$ and $\lambda =j(j+1)$

\end{document}